\documentclass[aps,prl,a4paper,floatfix,reprint,superscriptaddress]{revtex4-1} 
\usepackage{amsmath,amsthm,latexsym,amssymb,amsfonts,mathtools,epsfig}
\usepackage{graphicx,texdraw,natbib,subfigure,subeqnarray,fancyhdr,amsmath,multirow,color,bm,mathrsfs}
\usepackage[english]{babel}
\usepackage[utf8]{inputenc}			
\usepackage{overpic}
\usepackage{microtype}
\usepackage[colorlinks=true, citecolor=red, linkcolor=blue,urlcolor=blue ]{hyperref}
\newcommand{\integral}[4]{\int_{#1}^{#2}{#4}\mathrm{d}{#3}\,}

\newcommand{\BR}{\mathbf{R}}
\newcommand{\br}{\mathbf{r}}
\newcommand{\BD}{\mathbf{D}}

\begin{document}

\title{The non-Gaussian tops and tails of diffusing boomerangs} 

\author{Lyndon Koens}
\thanks{ML and LK contributed equally.}
\affiliation{Department of Applied Mathematics and Theoretical Physics, University of Cambridge, UK}

\author{Maciej Lisicki\normalfont\textsuperscript{*}}
\email{m.lisicki@damtp.cam.ac.uk}
\affiliation{Department of Applied Mathematics and Theoretical Physics, University of Cambridge, UK}
\affiliation{Institute of Theoretical Physics, Faculty of Physics, University of Warsaw, Poland}

\author{Eric Lauga}
\email{e.lauga@damtp.cam.ac.uk}
\affiliation{Department of Applied Mathematics and Theoretical Physics, University of Cambridge, UK}
\date{\today}

\begin{abstract}
Experiments involving the two-dimensional passive diffusion of colloidal boomerangs tracked off their centre of mobility have shown striking non-Gaussian tails in their probability distribution function  [Chakrabarty et al., Soft Matter {\bf 12}, 4318 (2016)]. This in turn can lead to anomalous diffusion characteristics, including mean drift. In this paper, we develop a general theoretical explanation for these measurements. The idea  relies on calculating the 
two-dimensional probability densities at the centre of mobility of the particle, where all distributions are Gaussian, and then transforming them to a different reference point. Our model clearly captures the experimental results, without any fitting parameters, and demonstrates that the one-dimensional probability distributions may also exhibit strongly non-Gaussian tops. These results indicate that the choice of tracking point can cause a considerable departure from Gaussian statistics, potentially causing some common modelling techniques to fail.
 \end{abstract} 
\maketitle

The spontaneous thermal agitation of small particles, called Brownian motion, was first observed by a famous botanist, Robert Brown, in grains of pollen in 1828 \cite{brown1828xxvii}. Since then, it has been recognised as a fundamental physical process with applications in many fields of biology \cite{Young2006,Lauga2016}, chemistry \cite{Doi1988a}, and physics \cite{Lisicki2014,Mo2015,Lisicki2016a}. In the early 1900s the nature of these agitations were characterised for spheres both  theoretically \cite{Einstein1905,Smoluchowski1906,Langevin1908} and experimentally \cite{Perrin1909}, thereby relating the internal micro-structure of a fluid to its macroscopic transport properties. This then allowed theorists to consider more complex systems, like the diffusion of colloids with non-spherical shapes \cite{Perrin1934,Brenner1967} or the ballistic behaviour of a particle shortly after agitation \cite{Hinch1975}. Only relatively recently have experiments managed to probe these non-spherical \cite{Han2006,Butenko2012,Kraft2013,Chakrabarty2014} and ballistic regimes \cite{Huang2011}. The ability to probe anisotropic shapes has in turn revealed exciting new behaviours \cite{Chakrabarty2013a,Chakrabarty2016}, prompting new theoretical models to explain them \cite{Koens2014, Cichocki2015, Cichocki2016, Lisicki2016}.

For an arbitrarily shaped {three-dimensional  particle moving in a Stokes flow}, there exists a special point, called the centre of mobility ({CoM}), at which the translation-rotation coupling mobility tensors are symmetric \cite{Brenner1967}. {This point can be found explicitly, given the mobility matrix at any point of the particle, using the transformation rules given explicitly in Ref.~\cite{Kim}.}  At the CoM, the {full} probability density functions (pdfs) remain Gaussian at all times, in agreement with the classical arguments of Brownian motion \cite{Cichocki2015}. In contrast, off this point the pdfs are not guaranteed to have the same statistics, {with both the mean and mean-squared displacement demonstrating transient behaviour not found at the CoM \cite{Chakrabarty2013a,Cichocki2015,Cichocki2012,Han2006}. These transient effects decay with the rotational time scales of the system, and the long-time limit diffusion rates are not only independent of position but also identical to those obtained at the  CoM \cite{Cichocki2012}.}

{In two dimensions, where only one rotational degree of freedom is present, the Stokes mobility matrix $\mathbf{M}$ becomes a $3\times 3$ tensor, composed of a $2\times 2$ translational part, a single rotational coefficient, and two coefficients coupling the rotational to translational motion. In this case an analogue of centre of mobility can be defined as the point where the translation-rotation coupling tensors vanish and in effect there is no coupling between translations and rotation \citep{Chakrabarty2014}. For a given two-dimensional mobility matrix the position of the analogue of the centre of mobility, which we will refer to as the two-dimensional centre of hydrodynamics (CoH), is determined by
\begin{equation}
\mathbf{r} = - \frac{1}{M_{\theta}}\left(M_{2\theta} \mathbf{\hat{x}}_{1}- M_{1\theta} \mathbf{\hat{x}}_{2}\right),
\end{equation}
where $\mathbf{r}$ is a vector from the frame origin to the CoH, $M_{\theta}$ is the rotational mobility coefficient, $M_{i\theta}$ is the coupling coefficient between rotation in the third direction and the spacial direction $i$ $(i=1,2)$ while $\mathbf{\hat{x}}_{i}$ denotes the unit vector in the $i$th spacial direction. Note the diffusion matrix $\mathbf{D}$ at any point is proportional to the mobility matrix via the fluctuation-dissipation theorem, $\mathbf{D}=k_B T \mathbf{M}$, where $k_{B}$ is the Boltzmann constant and $T$ is the temperature.
From the above equation, the corresponding diffusion matrix at the CoH can be determined using the standard two-dimensional mobility matrix transformation rules for Stokes flows \cite{Kim}. Physically, the two-dimensional centre of hydrodynamics plays an identical role to the three-dimensional CoM, in that the full pdfs determined by tracking this point remains Gaussian at all times. Furthermore, other points will, again, not necessarily generate the same statistics.}

\begin{figure}[b]
\begin{center}
\includegraphics[width=0.4\textwidth]{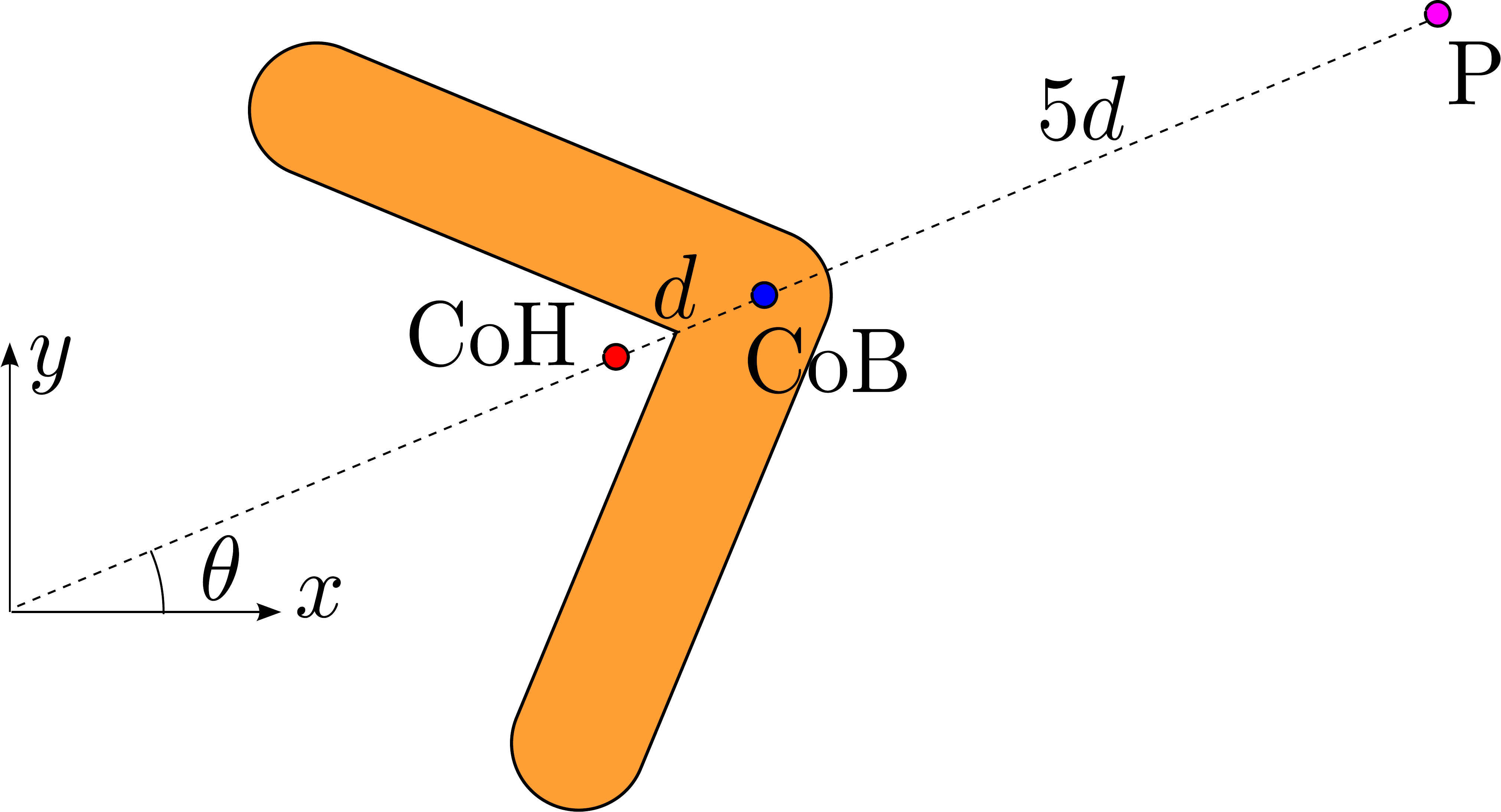}
\caption{Sketch of the boomerang particle used in the experiments of Ref.~\cite{Chakrabarty2016}, {where we indicate the location of} the reference point P, the geometric centre of the body (CoB) and the centre of hydrodynamics (CoH).}\label{boompart}
\end{center}
\end{figure}

{In an ideal world, experiments would only track the CoM and the CoH. However, this is often impractical, as these points could} lie off the body completely. Therefore one normally tracks a characteristic physical point of the particle, like the geometric centre, {which will experience transient, and possibly non-Gaussian statistics. In order to characterise these statistics Chakrabarty {\it et al.} considered the two-dimensional Brownian motion of a boomerang-shaped particle \cite{Chakrabarty2013a,Chakrabarty2014}. These shapes are useful as the CoH of these particles lies somewhere between their two arms, and can be displaced by varying the asymmetry in the boomerang arm lengths \cite{Chakrabarty2014}. In order to determine the position and orientation of the particles, high-precision tracking algorithms have been developed \cite{Chakrabarty2013}.  The works \cite{Chakrabarty2013a,Chakrabarty2014}
experimentally showed that the  mean and mean-squared displacement of a geometric point exhibited a crossover from short-time faster to long-time slower diffusion with the short-time diffusion coefficients dependent on the points used for tracking. This was in turn explained theoretically by solving a set of Langevin equations for the dynamics of the moments.} {Though these papers fully explained the dynamics of the mean and mean-squared displacement,  
 they did not characterise how non-Gaussian the pdfs at the geometric point were. Chakrabarty {\it et al.}~\cite{Chakrabarty2016} therefore aimed to characterise the two-dimensional (2D) Brownian motion of a boomerang-like particle when tracked off the CoH and to relate it to the diffusive properties of the particle. They concluded that the pdfs for the geometric centre (CoB, Fig.~\ref{boompart}) of the body exhibit strongly non-Gaussian tails when no initial orientation is imposed. These tails are not present when tracking the CoH. Qualitative arguments presented therein related the observations to the previously analysed general concepts of Brownian and non-Gaussian diffusion \cite{Wang2012}. However, the non-Gaussian behaviour was characterised  by fitting empirical relations to the measured distributions.}

In this paper, we provide a {quantitative} theoretical description for the pdfs observed by Chakrabarty~{\it et al.}~\cite{Chakrabarty2016}. The  integrals we obtained  are evaluated numerically before providing an analytical expression for the case when the drag is isotropic. Both these results replicate the experimental pdfs with no free parameters. Integrating out one of the spacial dimensions, we also show that these pdfs exhibit highly non-Gaussian configurations, even close to its mean value. Our results  emphasises that Gaussian statistics do not apply when tracking off the CoH. Furthermore, the  procedure highlighted in the paper, and the results therein, are generally {applicable} to any particle undergoing 
two-dimensional Brownian motions.
 
In two dimensions, {the configuration of a particle}  is described by three spatial variables: two describing position, $(x,y)$, and one describing orientation, $\theta$ (Fig. \ref{boompart}) {which measures the angle between the boomerang bisector angle and the $x$ axis. The coordinates are chosen in the laboratory frame in such a way that at $t=0$ the CoH is located at the origin.} The {complete} mobility tensor is thus a positive-definite  $3\times3$ matrix, which can be decomposed into the $2\times 2$ translational diffusion tensor, the rotational diffusion coefficient, and two off-diagonal $1\times 2$ and $2\times 1$ coupling sub-matrices.

The translational diffusion matrix in the frame of the particle can always be written {in a diagonal form with two coefficients only,}
\begin{equation}
\mathbf{D}=\begin{pmatrix}
D_{11} & 0 \\
0 & D_{22}
\end{pmatrix},
\end{equation}
while rotations are characterised by a single {rotational diffusion} coefficient, $D_\theta$.  The laboratory frame  diffusion matrix (denoted $xy$) is then given by $\BD_{xy}= \BR_\theta\cdot\BD\cdot\BR_\theta^{\mathsf{T}}$, where $\BR_\theta$ denotes the two-dimensional rotation matrix of angle $\theta$. Assuming that the particle's {motion is purely diffusive and that it} is initially located at the origin with zero deflection, the Gaussian probability distribution for the position of the particle $\br=(x,y)$, in the CoH, reads
 \begin{equation}
 p_{xy}(x,y,\theta;t) = \frac{1}{4 \pi \sqrt{D_{11}D_{22}}t } \exp\left(-\frac{\br\cdot\BD^{-1}_{xy}\cdot\br}{4 t}  \right),
 \end{equation}
{which is the classic solution to the Smoluchowski diffusion equation with diffusion matrix $\BD_{xy}$ \cite{VanKampen}.}

{We now turn to the angular probability distribution. Typically this distribution is required to be periodic with $\theta' \in (-\pi,\pi)$ and so should be represented by a so-called wrapped normal distribution \cite{Mardia}, 
\begin{equation} \label{wrap}
p_{\textrm{wrap}}(\theta',t)=\frac{1}{\sqrt{4 \pi D_\theta t}} \sum_{n=-\infty}^{\infty} \exp\left(-\frac{(\theta'+2 n\pi)^2}{4 D_\theta t}\right),
\end{equation}
which can also be formulated in terms of the Jacobi theta function of the third kind \cite{Jacobi}. However, as $\theta$ only appears within trigonometric functions outside the angular probability distribution function and we plan to integrate over it, we can define it to range from $-\infty$ to $\infty$ without loss of generality. Then, since there are no external torques acting on the particle, the Smoluchowski diffusion equation for the orientation angle yields a Gaussian distribution}
\begin{equation}\label{thetapdf}
p_{\theta}(\theta;t) = \frac{1}{\sqrt{4 \pi D_\theta t}} \exp\left(-\frac{\theta^2}{4 D_\theta t}  \right).
\end{equation}
{From inspection of Eqs.~\eqref{wrap} and \eqref{thetapdf}, it is obvious how the two probabilities are related. Both these distributions assume an initial orientation of 0; for a general initial angle $\alpha$, the argument of the angular distribution needs be replaced $\theta\to\theta-\alpha$.}

The complete probability distribution function (pdf) at CoH thus reads
\begin{equation}\label{pdfCoH}
P(x,y,\theta;t) = p_{xy}(x,y,\theta;t) p_\theta(\theta;t).
\end{equation}
The 2D pdf for the position at the fixed initial angle is then recovered by integrating out the angular degree of freedom as
\begin{equation}
P_{\mathrm{CoH}}(x,y;t) = \integral{-\infty}{\infty}{\theta}{P(x,y,\theta;t)}. \label{COHint}
\end{equation}
{Note that given the purely trigonometric dependence on $\theta$ in $p_{xy}$, the above integral would obviously be identical if the wrapped angular distribution was used instead of the Gaussian distribution since the wrapped integral is equivalent to dividing the infinite integral into $2\pi$ sections and then summing over all the respective parts.}

This integral generates Gaussian distributions with transient non-Gaussian tails \cite{Han2006}. These tails depend on $D_{11}-D_{22}$ and decay with $1/D_{\theta}$. If further averaged over all initial angles, however, these non-Gaussian tails disappear, in agreement with classical diffusion arguments. 

This picture changes significantly when a different tracking point is chosen. The pdf with respect to a different reference point, $\mathrm{T}=(x',y')$, is found by writing the coordinates of this point with respect to the centre of hydrodynamics and inserting them into Eq.~\eqref{COHint}. For Chakrabarty~{\it et al.}'s boomerang \cite{Chakrabarty2016} the coordinates become
\begin{align}
x' &=  x + \ell \sin\theta, \\
y' &= y + \ell \cos\theta,
\end{align}
where $\ell=d$ for the geometric centre of the body (CoB). {For the purpose of demonstration, following Chakrabarty {\it et al.}, we choose a more distant point $\rm P$ with $\ell=6d$.} The 2D pdf, with respect to the tracking point T, therefore becomes 
\begin{equation}\label{pdfT}
P_{\mathrm{T}}(x',y';t) = \integral{-\infty}{\infty}{\theta}{P(x' - \ell\sin\theta,y'-\ell\cos\theta,\theta;t)}.
\end{equation}

\begin{figure*}[t!]
\begin{center}
\includegraphics[width=0.75\textwidth]{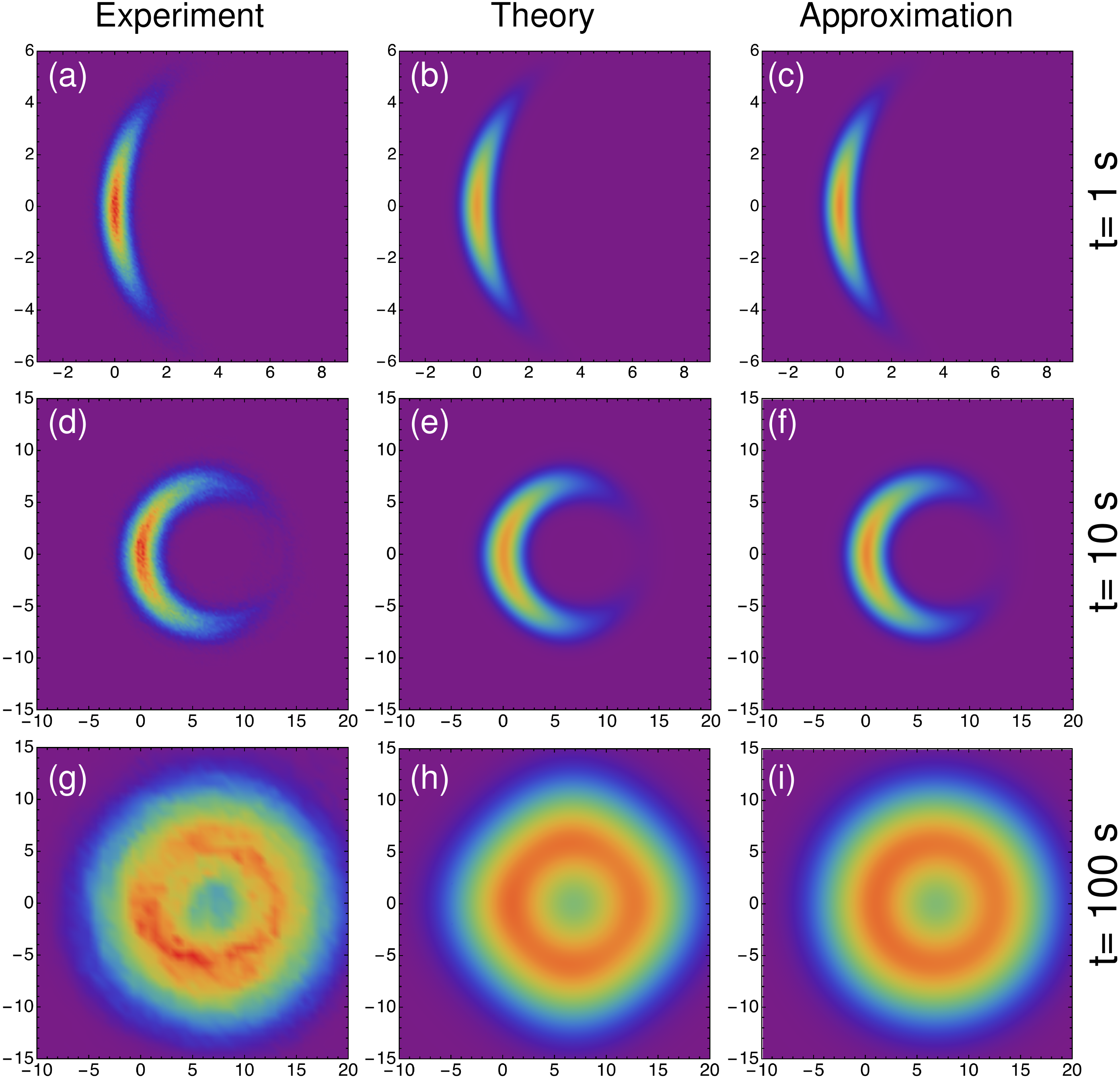}
\caption{Two-dimensional probability distribution function, $P(x,y,t)$, at the point P located  outside the boomerang. The experimental data  (left), originally shown in \cite{Chakrabarty2016}, and provided by the authors to be re-plotted here. The theoretical predictions (middle) are obtained by numerical integration of Eq.~\eqref{pdfT}, while the theoretical approximation (right) are obtained by truncating the series expansion in Eq.~\eqref{series} after 10  terms. }\label{2dpdfs}
\end{center}
\end{figure*}

Equation~\eqref{pdfT} can be integrated numerically to yield the 2D pdf for an arbitrary tracking point along the particle axis, and therefore theoretically predict the experimental results.
We plot in Figure~\ref{2dpdfs}  the predictions of Eq.~\eqref{pdfT} (middle column) and experimental results of Chakrabarty {\it et al.} \cite{Chakrabarty2016} (left column) for the tracking point P. For these numerical results, the diffusion coefficients, $D_{11}$, $D_{22}$ and $D_{\theta}$, and $d$ were taken to be the experimentally determined values (0.049~$\mu \mbox{m}^{2} s^{-1}$, 0.060~$\mu \mbox{m}^{2} s^{-1}$, 0.045~rad$^{2} s^{-1}$ and 1.133~$\mu \mbox{m}$ respectively).  This figure shows that Eq.~\eqref{pdfT} quantitatively captures the experimental behaviour. The slight discrepancy between the peak values of the pdfs are probably related to the finite sampling errors within the experiment.  All other experimental points show similar agreement. Similarly, averaging these distributions over all initial angles captures the Gaussian and non-Gaussian behaviour of Chakrabarty {\it et al.}'s one-dimensional radial distribution \cite{Chakrabarty2016}(not shown).

In the limit of isotropic drag, $D_{11}=D_{22}=D_{i}$, Eq.~\eqref{pdfT} can be evaluated exactly. This limit can capture much of the guiding physics and is relevant to many systems, including the boomerang particles which were noted to behave almost isotropically with $D_{i}\approx 0.058$~$\mu \mbox{m}^{2} s^{-1}$ \cite{Chakrabarty2016}. In this limit, $P_{\mathrm{T}}(x',y';t)$ becomes
\begin{equation} \label{iso}
P_\mathrm{T} = \frac{\beta \sqrt{\kappa}}{\pi^{3/2}}\integral{-\infty}{\infty}{\theta}{ e^{- \beta (r^2 +\ell^2) + 2 \ell \beta r \sin(\theta+\phi) - \kappa \theta^2}},
\end{equation}
where $\beta = 1/4 D_{i} t$, $\kappa=1/4 D_\theta t$ and we have written $x$ and $y$ in polar coordinates ($x= r \cos\phi$, $y= r \sin\phi$). The sinusoidal term in Eq.~\eqref{iso} can be expanded into a set of Fourier modes using the Jacobi-Anger expansion,
\begin{equation}
e^{i z \cos\theta} = \sum_{n=-\infty}^{\infty} i^{n} J_{n}(z) e^{i n \theta},
\end{equation}
thereby reducing the integral to an infinite series as
\begin{align}  \label{series}
P_\mathrm{T} =&\frac{\beta }{\pi}  e^{- \beta (r^2 +\ell^2)} \Bigl[I_{0}(2 \ell \beta r)\Bigr. \\ \nonumber 
& \Bigl.+2 \sum_{n=1}^{\infty}  I_{n}(2 \ell \beta r) \cos\left(\frac{n}{2} (\pi-2 \phi)\right) e^{-\frac{n^2}{4 \kappa}} \Bigr]  ,
\end{align}
where  $J_{n}(z)$ is the Bessel function of the first kind of order $n$ and $I_{n}(z) = i^{n} J_{n}(-i z)$ is the order $n$ modified Bessel function of the first kind. Clearly, the above expansion converges quickly for $t\neq 0$ and remains normalised for any summation truncation with $n>0$. At long times, $t \rightarrow \infty$,  $I_{n}(2 \ell \beta r) \rightarrow \delta_{n 0}$ and so the system returns to a simple Gaussian, while at short times $t \to  0$, all the separate Fourier modes become equally important ($\kappa \to \infty$) and the probability becomes skewed to a delta function located at $r=\ell$, $\phi=\pi/2$. This indicates that the non-Gaussian behaviour is again a transient effect which decays with $1/D_{\theta}$, consistent with the experimental result \cite{Chakrabarty2016}. However, unlike the anisotropy effect, this non-Gaussian behaviour still occurs if $D_{11}-D_{22}=0$ and instead depends critically on the value of $\ell$.

In Figure~\ref{2dpdfs} (right column) we plot the prediction from Eq.~\eqref{series} for the point P using $D_{i}=0.058$~$\mu \mbox{m}^{2} s^{-1}$. In each case the infinite summation was truncated at 10 terms. Again, a similar agreement is found for all the experimentally tracked points. For long times, the leading-order term is enough to reproduce the observed behaviour, while the stronger anisotropy at short times typically requires more terms.

{In order to} further demonstrate and quantify the non-Gaussian nature of these intermediate regimes, it is best to consider the one-dimensional distribution. This distribution can be obtained theoretically by further integrating Eq.~\eqref{pdfT} over one of the spatial dimension. Specifically we choose to integrate out $y$ to obtain a pdf which symmetric in $x$ for all times, ie.
\begin{equation} \label{1d}
P_\mathrm{T,1}(x;t) = \integral{-\infty}{\infty}{y}{P_\mathrm{T}(x,y;t)}.
\end{equation}
Experimentally this is equivalent to constructing a pdf from the laboratory $x$ positions for a given initial angle of $\theta=0$. Note this is different from the one-dimensional radial distribution used by Chakrabarty {\it et al.}~which averaged over all initial angles and observed a Gaussian core with non-Gaussian tails \cite{Chakrabarty2016}.
 
\begin{figure}[t]
\begin{center}
\includegraphics[width=0.42\textwidth]{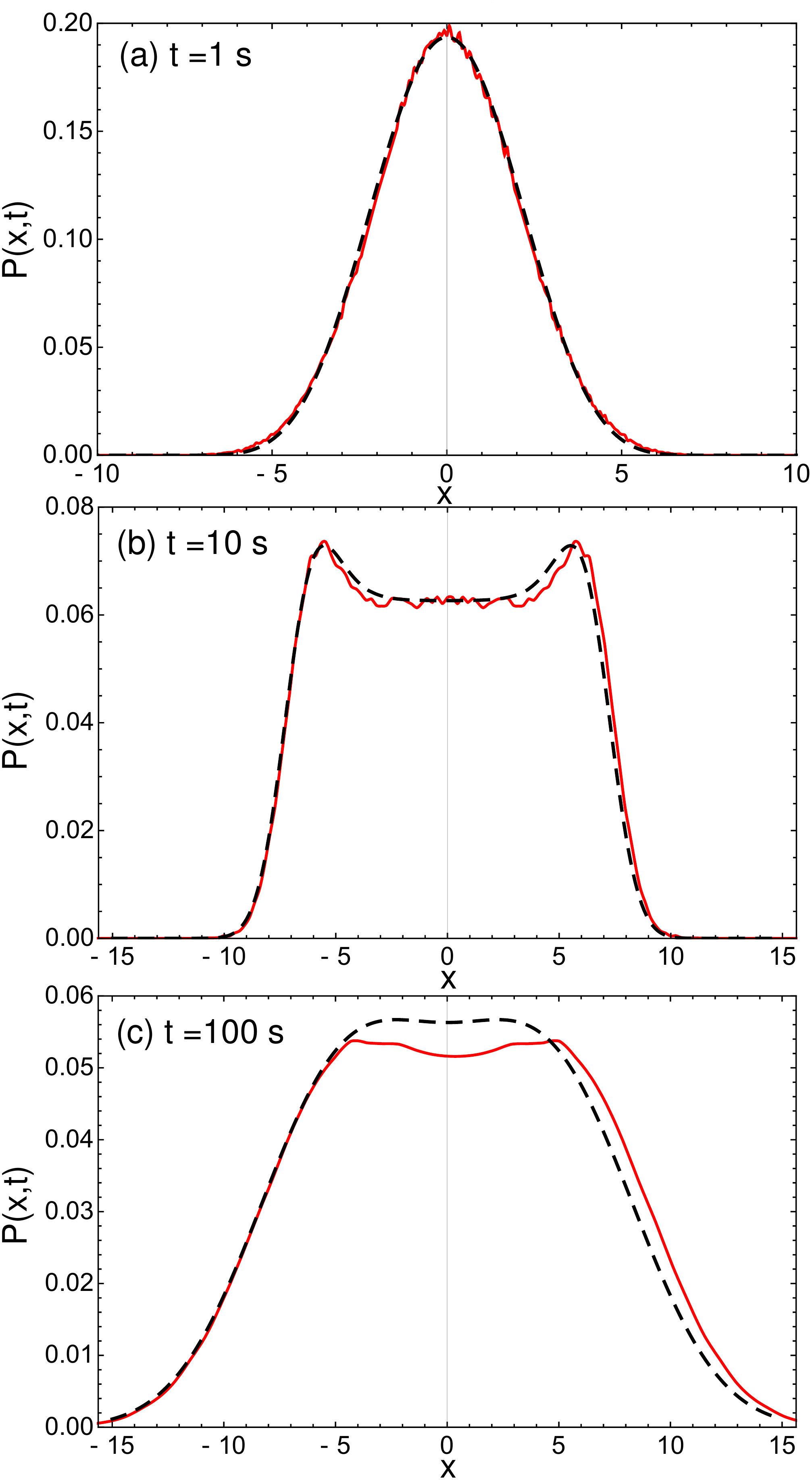}
\caption{One-dimensional pdfs $P(x,t)$ for the point P calculated by integrating out one dimension from the 2D pdfs from Eq.~\eqref{pdfT}, and plotted in black dashed lines, show strongly non-Gaussian tops. Red solid lines have been determined using the experimental data from Ref. \cite{Chakrabarty2016}.}\label{1dpdfs}
\end{center}
\end{figure}

 We display in 
 Figure~\ref{1dpdfs} the theoretical (dashed black lines) and experimental (solid red lines) 
 one-dimensional distributions for the point $\rm P$.  Theoretical dashed curves have been obtained by integrating out the $y$-dimension, as in Eq.~\eqref{1d}.  The solid red lines have been obtained by numerically integrating the 2D experimental data from Fig. \ref{2dpdfs}. The discrepancy between theory and experiment is again probably arising from experimental sampling limitations. Initially, the distribution shows a Gaussian like configuration which ultimately returns to a  Gaussian as $t \to \infty$. However, at intermediate times both the experiment and theory predict a highly non-Gaussian shape with two peaks. This multiply peaked structure reinforces the result that Gaussian statistics do not apply when tracking a particle off the CoH. This is especially true around the mean value of the system, where the central limit theorem would traditionally ensure Gaussian-like behaviour. This break down occurs because, when tracking a point off the CoH, the jumps in position are not independent, identically distributed random variables but are correlated with a `hidden variable', $\theta$. Therefore the sum of these jumps do not necessarily have to follow the central limit theorem, and so Gaussian statistics do not necessarily follow. 
 
 This is particularly relevant for many experimental and theoretical models which inherently assume that the pdf is roughly Gaussian. The Fokker-Plank and Langevin equations are two such examples, both of which assume that the system is described sufficiently by the first two moments of the fluctuations, i.e.~a Gaussian. These models typically work well for the behaviour of the CoH or when the full particle configuration is being resolved. However, as Fig.~\ref{1dpdfs} shows, the Gaussian assumption breaks down when the dynamics of an arbitrary point with marginalised configuration dimensions is analysed. Therefore, in these cases it is inappropriate to write down the equations in their traditional form. Rather, when the behaviour of a point other than the CoH is desired, it is best to use models that do not assume Gaussian behaviour, like the Master equation \cite{VanKampen}, or to transform the equations from the CoH to the relevant point before or after they are solved.

\begin{figure*}[t]
\begin{center}
\includegraphics[width=0.8\textwidth]{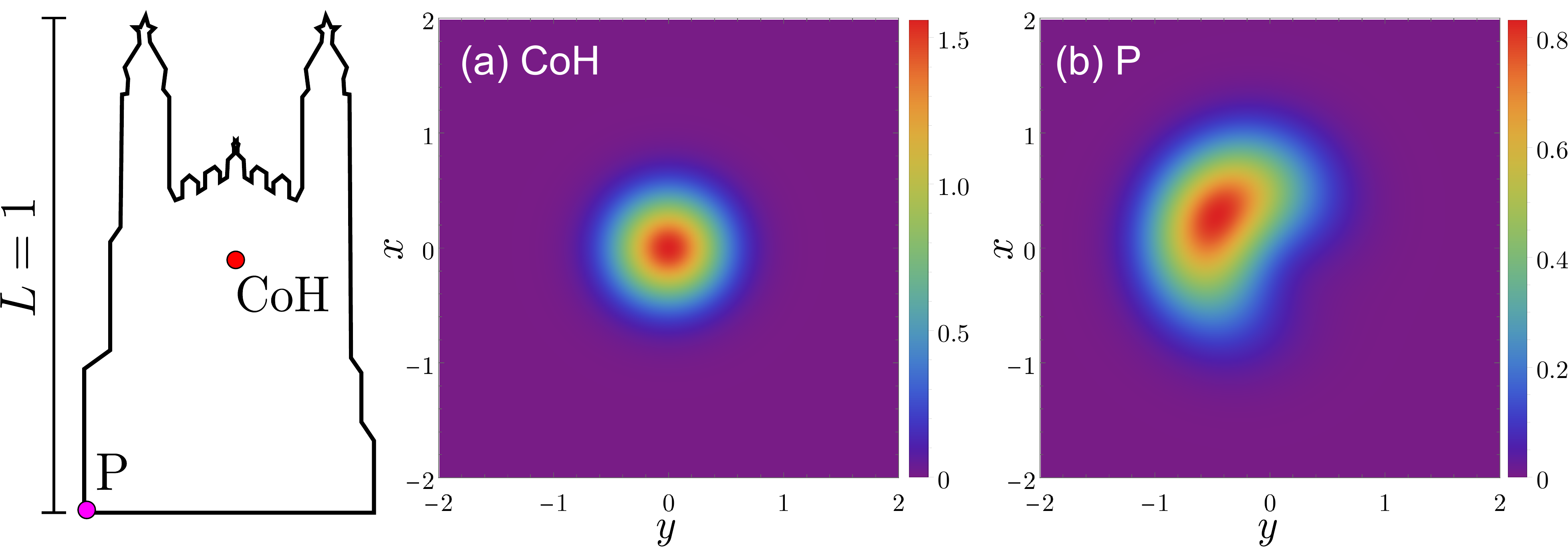}
\caption{Two-dimensional pdfs $P(x,y,t)$ for the diffusing King's College Micro-Chapel silhouette (sketched), obtained when tracking (a) the centre of hydrodynamics CoH  and (b) the corner P of the particle at an early time $t=0.3$.}\label{kings}
\end{center}
\end{figure*}

The  results derived in this paper can be applied to colloids of any shape, provided its 2D translational and rotational mobility matrix is known. In order to illustrate the application to a more complex shape, we consider a diffusing silhouette of a Cambridge landmark -  the King's College Chapel \cite{Kings}. The shape is constructed out of rod-like segments, as shown in Fig.~\ref{kings}, and the total mobility matrix is then calculated using resistive force theory with the assumptions for the drag coefficients perpendicular and parallel to a unit segment to be $\zeta_{\perp}=2\zeta_{\parallel}=2$ \cite{GRAY1955,Lauga2009}. 
If the centre of mobility is chosen as reference point, the procedure outlined in our note leads to the expected Gaussian distribution (Fig. \ref{kings}a). However, when a more convenient tracking point is chosen, such as the corner P, the resulting distributions are inherently non-Gaussian (Fig. \ref{kings}b).

We further remark that in three dimensions the mobility matrix, measured from the centre of mobility, will have non-vanishing coupling components if the particle is chiral. This means that translation and rotation cannot be decoupled and so the marginal pdfs may not be Gaussian, even at short times, although the underlying full-dimensional (spatial and orientational) pdf at CoH would be Gaussian at all times. 
To understand this phenomenon, a more general analysis will be needed.

In summary, classical Brownian motion arguments accurately describe the particle's dynamics when tracking the centre of mobility (CoH). Recently, however, two-dimensional experiments have shown that anomalous diffusion occurs when tracking a different point, generating mean displacements and non-Gaussian tails in the particles probability distribution function \cite{Chakrabarty2016}. In this paper, we developed a general theoretical procedure to explain the non-Gaussian effects seen by the experiments. This method can be solved either numerically or analytically in the case of isotropic drag. Using the mobility matrix reported in Ref.~\cite{Chakrabarty2016}, both methods quantitatively captured the non-Gaussian experimental results without any {additional free} parameters.  Similarly to the experiment, we observed that this non-Gaussian behaviour is transient, decaying with $1/D_{\theta}$. However, further exploration of the experimental and theoretical results revealed that in addition to the non-Gaussian tails seen previously, the one-dimensional pdf (defined by Eq.~\ref{1d}) has highly non-Gaussian behaviour near its mean. This occurs because the orientation angle $\theta$ is correlated to the positional jumps when off the CoH, thereby violating the central limit theorem assumptions of independent identically distributed random variables. These correlations may render it inappropriate to use the Fokker-Planck formulation for an arbitrary point. The results in this paper are thus very general and can be applied to any two-dimensional diffusing particle {with known translational and rotational diffusion coefficients, either taken from experiments, or computed using a variety of numerical methods \cite{GRAY1955,Lighthill1975,Johnson1980,Carrasco1999,Torre2002,EkielJezewska2009}}
such as a silhouette of King's College Chapel (Fig.~\ref{kings}) or even any useful shape that does not look like a famous  landmark.

\section*{Acknowledgements}
The authors thank Qi-Huo Wei and Ayan Chakrabarty for sharing their experimental results and helpful discussions about their measurement procedures. This research was funded in part by an ERC grant to EL and a Mobility Plus Fellowship from the Polish Ministry of Science and Higher Education to ML.

\bibliographystyle{phaip}

\begin{thebibliography}{10}
\bibitem{Jacobi}
{The Jacobi theta function is defined as
  $\Theta(\theta,\tau)=\sum_{n=-\infty}^{\infty} w^{2n} q^{n^2}$, with
  $w=\exp(i\pi\theta)$ and $q=\exp(i\pi\tau)$, and thus the wrapped
  distribution \eqref{wrap} can be written as $(2\pi)^{-1}\Theta(\theta/2\pi, i
  D_\theta / \pi)$. Noticeably, the Gaussian distribution defined from
  $-\infty$ to $\infty$ yields the same average values as the wrapped
  distributions when the integrand involves only trigonometric functions.}
\end{thebibliography}

\end{document}